
\input phyzzx

\REF\cghs{C.~G.~Callan, S.~B.~Giddings, J.~A.~Harvey and
A.~Strominger
\journal Phys. Rev. & D45 (92) R1005.}

\REF\hawi{S.~W.~ Hawking
\journal Comm .Math. Phys. & 43 (75) 199.}

\REF\polya{A.~M.~ Polyakov
\journal Phys. Lett. & B103 (81) 207.}

\REF\gid{S.~B.~Giddings, {\it Toy Models for Black Hole Evaporation},
UCSBTH-92-36,
 June 1992 and references therein.}

\REF\hs{J.~A.~Harvey, A.~Strominger, {\it Quantum Aspects of
Black Hole}, EFI-92-41, September 1992, and references therein.}

\REF\mty{M.~S.~Morris, K.~S.~Thorne and U.~Yurtsever
\journal Phys. Rev. Lett. & 61 (88) 1446.}

\REF\fmnekty{J.~Friedman, M.~S.~Morris, I.~D.~Novikov, F.~Echeverria,
G.~Klinkhammer, K.~S.~Thorne, U.~Yurtsever
\journal Phys. Rev. &D42 (90) 1915.}

\REF\fro{V.~P.~Frolov
\journal Phys. Rev. &D43 (91) 3878.}

\REF\kith{W.~ Kim, K.~S.~ Thorne
\journal Phys. Rev. & D43 (91) 3929.}

\REF\hawii{S.~W.~ Hawking
\journal Phys. Rev. & D46 (92) 603.}

\REF\kay{B.~S.~Kay, {\it The Principle of Locality and Quantum Field
Theory on (Non Globally Hyperbolic) Curved Spacetimes}, DAMTP/R-92/22,
July 1992.}

\REF\hael{S.~W.~ Hawking, G.~F.~R.~ Ellis, {\it The Large Scale
Structure of Space-Time}, Cambridge Univ. Press, (1982).}

\REF\rsti{J.~G.~Russo, L.~Susskind, L.~Thorlacius
\journal Phys. Rev. & D46 (92) 3444.}

\REF\sak{N.~ Sakai, {\it c=1 Two Dimensional Quantum Gravity},
TIT/HEP-205, \hfill \break
 August 1992, \nextline
Y.~ Matsumura, N.~ Sakai, Y.~ Tanii, T.~ Uchino,  {\it Correlation
Functions in Two Dimensional Dilaton Gravity},
TIT/HEP-204, STUPP-92-131, August 1992.}

\REF\yos{M.~Yoshimura, {\it Quantum Decay of de Sitter Universe and
Inflation without Fine Tuning in 1 + 1 Dimensions},
TU/92/416, September 1992, \nextline
M.~Hotta, Y.~Suzuki, Y.~Tamiya and M.~Yoshimura, {\it Quantum Back
Reaction in Integrable Cosmology},
TU/93/426, January 1993.}

\def\p{\partial}
\def\phip{\p_+\phi}
\def\phim{\p_-\phi}
\def\phipm{\p_+\p_-\phi}

\def\rhopm{\p_+\p_-\rho}

\def\xp{x^+}
\def\xm{x^-}
\def\up{u_+}
\def\um{u_-}
\def\xchon{x^{\prime}}
\pubnum={218/COSMO$-$29}
\titlepage
\vskip 1.5cm
\title{Chronology Protection in Two-Dimensional Dilaton Gravity}
\vskip 2.0cm
\author{ Takashi Mishima \footnote{\dagger}{On leave of absence from
 Institute for Nuclear Study,
    University of Tokyo,
     Midori-cho, Tanashi-shi, Tokyo 188, Japan}
      \ and  \ Akika Nakamichi \footnote{\ast}{e-mail address:
       akika@phys.titech.ac.jp }
}
\vskip 1.5cm
\address{
${\dagger}$
Physical Science Laboratories, College of Science and Technology,
 \nextline
Nihon University, Narashinodai, Funabashi, Chiba 274, Japan
 \nextline
 \nextline
 ${\ast}$ Department of Physics,
    Tokyo Institute of Technology
    \nextline Oh-okayama, Meguro-ku, Tokyo 152, Japan
        }
\vskip 1.0cm
\abstract{The global structure of 1 + 1 dimensional compact
Universe is studied in two-dimensional model of dilaton gravity.
 First we give a classical solution corresponding to the spacetime
in which a closed time-like curve appears, and show the
instability of this spacetime due to the existence of matters.
 We also observe quantum version of such a
spacetime having closed timelike curves never reappear
unless the parameters are fine-tuned.
 }
\vfill
\eject
	Callan, Giddings, Harvey and Strominger[\cghs]
(abbreviated as CGHS)
provided a useful toy model of (1+1)-dimensional gravity.
They attempted to
analyze the back-reaction of the Hawking radiation[\hawi] in the
two-dimensional analogue of black hole geometry
in a consistent way by the use of this model.
In their first paper, CGHS developed their original scenario as follows:
First introducing 1+1 dimensional dilaton gravity, CGHS found the
solutions corresponding to black
hole formation, and showed occurrence of the Hawking radiation in the
spacetime background they obtained. Next taking account of quantum effect  of
quantized conformal matters as the Polyakov term[\polya],
they analyzed the back-reaction of the Hawking effect in a leading
semi-classical approximation consistently.
However their first scenario on the quantum black hole remains to be
elaborated. Up to the present time, many people have studied the behavior
 of quantum black hole in this model[\gid,\hs].

	On the other hand, in general relativity, there are other
interesting problems which must be resolved taking account of quantum
effects.
For example, since Morris, Thorne and
Yurtsever pointed out the possibility of making a time-machine[\mty],
quantum effect on the spacetime in which closed time-like curves appear
(abbreviated as CTC-spacetime) has been discussed by several authors
[\fmnekty-\kay]. One of the most important problems in this subject
is whether CTC-spacetime continues to exist or not
if the quantum effect is taken into account. Hawking suggested the
possibility that the laws of physics may prevent the appearance
of closed time-like curves
(so called `Chronology Protection Conjecture'). In four dimensions the
analysis of this problem so far have been done only in fixed background
spacetimes. So it will be interesting to treat such spacetimes including
back-reaction  in closed form even in lower dimensions.

	In this paper we will investigate the quantum back-reaction
problem on the stability of the CTC-spacetime in two-dimensional model
of dilaton gravity. Following the CGHS scenario, we apply this model to
a compact one dimensional universe.
In the first half of this paper we give a classical solution
corresponding to CTC-spacetime: an analogue of the Misner universe.
Then we show the disappearance of this spacetime due to the existence
of conformal matters even if the parameters are fine-tuned.
In the second half we investigate whether the extension to
such a CTC-spacetime can be made or not if quantum back-reaction is
taken into account.

	 We consider the $ 1+1 $ dimensional renormalizable theory of
gravity coupled to a dilaton scalar field $\phi$ and $N$ massless
conformal  fields $ f_i$.
 The classical action is
$$
\eqalign{
  S
 = {1 \over 2\pi} \int d^2 x \sqrt{-g}
   \bigl[e^{-2\phi} (R + 4(\nabla \phi)^2 - 4\lambda^2)
-{1 \over 2} \sum^N_{i=1} (\nabla f_i)^2 \bigr]                   \cr
        } \, ,    \eqno\eq
$$
where $R$ is the scalar curvature, $\lambda^2$ is a cosmological
 constant.
This model differs from the original C.G.H.S. model in the sign of the
cosmological term.

The equations of motion derived from (1) are
$$
\eqalign{
 0
 =& - 4\phipm + 2\rhopm + 4\phip \phim - \lambda^2 e^{2\rho}   \,,     \cr
 0
 =& \phipm - \rhopm \,,                                             \cr
 0
 =& \p_{+}\p_{-} f_{i} \, ,                                          \cr
        }
\eqno\eq
$$
in the conformal gauge:
$
 g_{\mu\nu} dx^{\mu} dx^{\nu}
 = -e^{2\rho} dx^{+} dx^{-} \, ,
$
where $x^{\pm} = t \pm x$ . In addition the following constraints have
to be imposed:
$$
 4e^{-2\phi}(\p_{\pm}^{2}\phi -2\p_{\pm} \rho \p_{\pm} \phi)
 = \sum_{i=1}^N \p_{\pm}f_{i} \p_{\pm}f_{i} \, .
\eqno\eq
$$

 In the following we adopt the periodic boundary condition that the
spacetime point $(t,x)$ is identified with $(t,x+L)$ and the initial
condition that the Universe starts from a static cylinder spacetime endowed
with the usual Minkowski metric at the past infinity.
Then the general form of the solutions is given by:
$$
\eqalign{
e^{-2\phi}
 =& \up + \um + e^{-2\lambda t}   \, ,               \cr        e^{2\rho}
 =& e^{-2\lambda t}e^{2\phi}   \, ,                                   \cr
         }
\eqno\eq
$$
where $u_+$ and $u_-$ are chiral periodic functions which satisfy the
following equations from the constraints (3):
$$
 0
 = \p _\pm ^2 u_\pm + \lambda \p _\pm u_\pm
   + {1 \over 2} \p _\pm f \p _\pm f \, .
\eqno\eq
$$

	 If there is no matter field, general solutions  satisfying the
periodic boundary condition depend only on time:
$$
\eqalign{
 e^{2\phi}
 =& (M + e^{-2\lambda t})^{-1} \, ,                                   \cr
 e^{2\rho}
 =& e^{-2\lambda t}e^{2\phi}   \, ,                                   \cr
         }
\eqno\eq
$$
where $M$ is an arbitrary constant and corresponds to an initial value
for $\phi$ imposed at some past time. We classify the behavior of the
solutions into three types with respect to the sign of  $M$.

 	When $M$ equals to 0, the solution becomes an analogue of the Linear Dilaton
Vacuum solution in C.G.H.S. model.  The world is a static
cylinder spacetime.

  	When $M$ is negative, we see from (6) that the observer meet
some singularity in a finite proper time. From the expression of scalar
curvature $R$:
$$
 R
 = -{ 4\lambda^2 M \over M+e^{-2\lambda t} } \, ,
\eqno\eq
$$
we can see that this singularity is a true singularity.
In fact this singularity is the same as the one in the 1 + 1 dimensional
black-hole treated in CGHS.

	On the other hand when $M$ is positive, the space collapses
into zero volume in a finite proper time (as coordinate time
$t$ goes to $+\infty$). But from (7) the scalar curvature still remains
a finite value $-4\lambda^2$
at the point. Hence we expect that the spacetime can be extended.
In fact, if one defines the coordinates:
$$
\left \{ \matrix{
 \eta
 =& -e^{-2\lambda t}  \, ,                        \cr
 \psi
 =& t \pm x           \, ,                        \cr
                }
\right.
\eqno\eq
$$
the metric becomes
$$
 ds^2
 = {-1 \over M - \eta} (\eta d\psi^2 + {1 \over \lambda} d\psi d\eta) ,
\eqno\eq
$$
which is analytic in the extended manifold defined by
$\psi$ and by $-\infty < \eta < M$.

	The behavior of the extended manifold is shown in Fig.1.
The region $\eta < 0$ is isometric with the previous manifold.
The region $\eta > 0$ is extended part, where the closed timelike curves
appear, because the roles of $t$ and $x$ are interchanged.
It should be noted that the spacetime has a naked singularity on $\eta = M$
where the dilaton field becomes $+ \infty$.
The surface $\eta$ = 0 is the boundary of the Cauchy Development;
that is the Cauchy Horizon, where the dilaton remains finite.
This extension is achieved by the same way as in the case of Misner
space (we have two-dimensional version when $\lambda  = 0 $) and
the Taub-NUT space [\hael].
Hawking used the Misner space to discuss the chronology protection
conjecture [\hawii]. In the rest of the present paper we investigate
the conjecture in this model both at classical and the quantum levels.

	Next we consider the Universe with classical massless scalar
fields. The solutions of (3) satisfying the periodic boundary condition
always exist for an arbitrary configuration of the scalar fields.
 If we expand the scalar fields in Fourier series:
$$
\eqalign{
 f(\xp, \xm)
 =& f_+ (\xp) + f_- (\xm) \, ,              \cr
 f_{\pm}
 =& \alpha_{\pm} + \sqrt{{\varepsilon \over 2}} x^{\pm}
    + \sum^{\infty} _{n=1}\bigl( a^{\pm}_n \sin {2\pi n \over L} x^{\pm}
               + b^{\pm}_n \cos {2\pi n \over L} x^{\pm} \bigr) \, , \cr
        }
\eqno\eq
$$
where $\alpha_{\pm} , a^{\pm}_{n}$ and $b^{\pm}_{n}$ are the expansion
coefficients. We obtain the solution as follows
$$
\eqalign{
 e^{-2\phi}
 =& M
   - \bigl\{ {\varepsilon \over 2\lambda}
       + {1 \over 2\lambda} \Sigma \bigr\} t
   + e^{-2\lambda t}
   + \bigl(\, oscillation\  part \, \bigr)\, ,    \cr
 \Sigma
 =& \sum_{(\cdot)=\pm} \sum^{\infty} _{n=1}\,({2\pi n \over L})^2 \,
[(a^{(\cdot)}_{n}) ^{2}
                                   + (b^{(\cdot)}_n) ^2]\, ,          \cr
        }
\eqno\eq
$$
where the fourth term on the right hand side of the first equation in
(11) is the oscillation part, that is the sum of the trigonometric
functions.
  From (11) it should be noted that any classical configuration
of matter fields makes a finite contribution to the term proportional to
time, which causes divergence of the scalar curvature.
Therefore the Universe inevitably meets singularity at $t=\infty$ and
cannot extend to the region with closed time-like curves.

	From now on we study how the classical solution change
if we include back-reaction.
In two dimensions, the quantum effect of massless matter fields
is completely determined by conformal anomaly[\polya].
The quantum effective action is sum of the classical action (1) and
the Polyakov term induced by the $N$ matter fields:
$$
 S_{quantum}
  = - {\kappa \over 8\pi} \int d^2x\sqrt{-g(x)}
                          \int d^2{\xchon}\sqrt{-g(\xchon )}
              R(x) \, G(x, \, \xchon) \,R(\xchon)  \> ,
\eqno\eq
$$
where $\kappa$ is ${N\over 12}$ and  $\  G(x, \, \xchon)$ is a
Green's function of the scalar fields.
We assume that $\kappa$ is to be a large number and use the
$1/N$-expansion.
Further we add the following term introduced
by Russo et al.[\rsti] to the above action:
$$
 S_{quantum}^\prime
 = - {\kappa\over 8\pi} \int d^2x\>\sqrt{-g} 2\phi \,R \> .
\eqno\eq
$$

	Making the field redefinition:
$$
\eqalign{
  \chi
  =& \sqrt{\kappa}\bigl( \, \rho -{1 \over 2}\phi
                                 + {1 \over \kappa}e^{-2\phi}\,
                  \bigr)\, ,                                       \cr
  \Omega
  =& \sqrt{\kappa}\bigr( \, {1 \over 2}\phi +{1 \over \kappa}e^{-2\phi}\,
                 \bigr)\, ,                                       \cr
       }
\eqno\eq
$$
we see that the semi-classical equations of motion is simplified;
$$
\eqalign{
 &\partial_{+}\partial_{-} \chi
 = -{\lambda^2 \over \sqrt{\kappa}}
             \, e^{{2 \over \sqrt{\kappa}}(\chi - \omega)} \, , \cr
 &\partial_{+}\partial_{-}( \chi - \omega)
 = \, 0   \, ,                                                    \cr
        }
\eqno\eq
$$
and
$$
 \partial_{\pm}f \cdot \partial_{\pm}f -2\kappa t_{\pm}(x_{\pm})
 = -\partial_{\pm}\chi \partial_{\pm}\chi
   +  \partial_{\pm}\Omega \partial_{\pm}\Omega
   +\sqrt{\kappa} \partial^2 _{+}\chi \, ,
\eqno\eq
$$
where $\cdot$ denotes the sum over $i$ and $t_{\pm}$ are arbitrary
chiral functions to be determined by the boundary conditions. In this
case the classical matter fields $f_i$ are interpreted
as macroscopic behavior of quantized fields.
For an arbitrary macroscopic part of quantized matter fields as shown
in (10), the solution of the equations (15) and (16) is given by
$$
\eqalign{
 \sqrt{\kappa} \chi
  =& M - {2\kappa \over \lambda}
    \bigl( \, {\lambda^2 \over 4} - t_{vev}
    + {\varepsilon + \Sigma \over 4\kappa} \bigr) \, t
    + e^{-2\lambda t}
    + \bigl(\, oscillation\  part \, \bigr)\, ,   \cr
 \sqrt{\kappa} \Omega
  =& M + {2\kappa \over \lambda}
    \bigl( \, {\lambda^2 \over 4} + t_{vev}
    - {\varepsilon + \Sigma \over 4\kappa} \bigr) \, t
    + e^{-2\lambda t}
    + \bigl(\, oscillation\  part \, \bigr)\, ,                      \cr
        }
\eqno\eq
$$
where $t_{vev} \equiv t_{\pm}$ is determined to be ${\pi ^2 \over L^2}$
due to the Casimir effect, and $\Sigma$ is the same as in (11).

  	From the equations (14) and (17), we can see qualitative
behavior of $\phi$ as a function of time $t$, when the position $x$
is fixed (Fig.2).

  	We examine whether quantum version of CTC-spacetime realizes
or not.   From (14) and (17), we obtain $
2\rho
 = 2\phi - 2\lambda \, t .
$
To extend the spacetime
to the region with CTC's, $2 \rho $ must become linear in $t$
as $t \rightarrow \infty$ in the conformal flat gauge,
and also the coefficient of the linear term must be negative.
By comparing $\Omega \, - \, \phi$ and $\Omega \, - \, t$ relations,
we recognize that there are two distinct types of solutions.
 One is realized in the case (i): $ M >\, \sqrt{\kappa} \, \Omega_{cr}\,$
 in the following parameters:
$$
\eqalign{
&a^{(\pm)}_{n}
 =\, b^{(\pm)}_{n} = 0 \ \ \ \ \ \ \ \ \     \forall{n}\, ,  \cr
& \varepsilon
 =\, \kappa \lambda^2 + 4\kappa t_{vev}\, ,
       }
\eqno\eq
$$
where $\Omega_{cr}$ is the local minimum of $\Omega$ at $\phi = \phi_{cr}$
(Fig.2(a)), and the first condition in (18)
restrict the universe to a homogeneous one.
The other is realized in the case (ii): $ \Omega_{min} = \Omega_{cr} $
 ($\Omega_{min}$ is the local minimum of $\Omega$
in $\Omega \, - \, t$ relation (Fig.2(b))),
$ a^{(\pm)}_{n} =\, b^{(\pm)}_{n} = 0 \, (\forall{n})\, $ and some appropriate
negative $M$ is chosen. In the case (i), the value of dilaton at $\eta = 0$
can be adjusted to be so small that the semi-classical approximation is valid.
On the other hand in the case (ii), the spacetime can be extended over
$ \phi = \phi_{cr} $ to the strong coupling region smoothly and
the appearance of CTC-spacetime may occur in the limit of
$t = \infty$ where $\phi$ becomes infinite. However in the case (ii),
the analysis with full quantization is needed. The statement that
whether CTC's appear or not may becomes meaningless.

	Finally we conclude that the spacetime never been extended
across $\eta = 0$ with any configuration of quantized matters
except the fine-tuned example as in (18).
Thus it can be said that the chronology protection holds
in a $\it{weaker}$ sense in the semi-classical level including
back-reaction.

 	In order to determine whether the chronology protection holds or not
in a $\it{strong}$ sense, We must go on to extend
the  classical and semi-classical treatments of the compact universe
in this paper to the full quantization of two-dimensional dilaton
gravity (for example, see [\sak]). For the existence of any consistent
solution in the extended region ($\eta > 0$) inevitably depends on
the information from the naked singularity($\eta = M$), whose neighborhood
is very strong coupling region. To proceed the analysis, the construction of
the physical states having such classical and semi-classical behaviors
will be intriguing especially.

	Recently another interesting case has been reported in [\yos],
in which two-dimensional analogue of inflation is treated.
 CGHS scenario has been shown to be extensively useful for the study of
back-reaction problems in general relativity.

\nextline
	We are most grateful to A. Hosoya for a collaboration in an early stage of the
work and reading the manuscript. One of the authors(T. M.) would like to
acknowledge Y.~Onozawa, M.~Siino and K.~Watanabe for enjoyable discussions.
\refout
\vskip 2cm

\noindent {\bf Figure Captions}
\vskip 1.5cm
\item{\bf Fig. 1}Misner-type universe;
\nextline
$t=+\infty \, (\eta=0)$ is a closed null geodesics. The region:$(\eta<0)$ is
globally hyperbolic spacetime, and the region:$(\eta>0)$ have closed
time-like curves.
\vskip 1cm
\item{\bf Fig.2}$\Omega \,$-$\, \phi$ and $\Omega \,$-$\, t$ relations;
\nextline
(a) $\Omega \,$-$\, \phi$ relation, (b) $\Omega \,$-$\, t$ relation

\end